\documentclass[aps,pra,twocolumn,superscriptaddress,showpacs]{revtex4-2}
\usepackage{amsmath,amssymb,amsfonts}

\usepackage{graphicx}

\begin{document}

\title{Determination of Quantum Instrument Parameters\\for a Stern-Gerlach
Non-ideal Device}

\author{Ioan Sturzu}
\thanks{Present: isturzu@gmail.com. Original affiliation: ``Transilvania'' University, Department of Physics,
Eroilor 29, Bra\c{s}ov, Romania.}

\begin{abstract}
In the Stern--Gerlach (SG) apparatus, when the filtering is not perfect, the 
fermion state transformation is done by non-ideal devices. Such an apparatus
is described in quantum mechanics by quantum instruments. It will be shown 
how the parameters of this instrument can be expressed in terms of some 
experimentally accessible quantities. These relations are valid both for 
single and successive measurements. In the latter case, the measuring device 
is rotated with respect to the first one.
\end{abstract}


\maketitle

\section{Introduction}

The Operational Quantum Physics (OQP) is a formalism based on the concept of
quantum instrument (see e.g. \cite{Ludwig,Davies,Ludwig2}, \cite{Ludwig3},
\cite{Kraus1}, \cite{Martinez,Lahti,Busch,Busch1} and references therein)
and is very useful when the non-ideal measurement devices are used in 
experiments. In many cases, as in the Stern--Gerlach experiment, the filtering 
is not perfect, so the fermion state transformation cannot be described by 
an ideal projection-valued measure. The quantum instrument formalism gives a
better description of the measurement process (see, for example 
\cite{Ludwig2}, \cite{Schroeck}). Here one will use the OQP
formalism in order to identify the quantum instrument describing the
non-ideal measurements in a Stern--Gerlach apparatus.

The quantum instrument is a mathematical object describing the
measurement process. A short presentation of this concept will be given
here. Let us consider a compound system $S+B$ (system plus bath) which is
described in the Hilbert space ${\cal H}={\cal H}_{s}\otimes {\cal H}_{b}$. 
After the preparation procedure, the system is in a mixed state described by
the statistical operators $\hat{\rho}_{s}$ and $\hat{\rho}_{b}$
respectively, and the total state $\hat{\rho}(t_{0})$ at $t=t_{0}$ 
can be written in a factorized form:
\[
\hat{\rho}(t_{0})=\hat{\rho}_{s}(t_{0})\otimes \hat{\rho}_{b}(t_{0})
\]%
Consider that the measurement starts at the subsequent moment $t_{i}$ and
and ends at the moment $t_{f},$ so the compound system Hamiltonian is:
\[
\widehat{H}\left( t\right) =
\left\{
\begin{array}{ll}
\widehat{H}_{0}, & t < t_{i},\quad t > t_{f} \\
\widehat{H}_{0}+\widehat{H}_{\text{int}}(t), & t \in [t_{i}, t_{f}]
\end{array}
\right.
\]

The compound system state for $t>t_{f}$ is given by the unitary evolution:
\[
\hat{\rho}(t)=\hat{U}(t,t_{0})\hat{\rho}_{s}(t_{0})\otimes \hat{\rho}%
_{b}(t_{0})\hat{U}^{\dagger }(t,t_{0})
\]%
\[
\hat{U}(t,t_{0})=\mathbb{T}\exp \left( -\frac{i}{\hbar }\int_{t_{0}}^{t}dt^{%
\prime }\widehat{H}\left( t^{\prime }\right) \right)
\]%
The result of this operation is that the state $\hat{\rho}_{s}(t)$ is not
anymore pure.

Generally, any mixed state can be decompose in a convex combination of pure
statistical operators. An orthogonal decomposition is given by the spectral
theorem for the selfadjoit operator which is $\hat{\rho}_{s}(t)$. Apart from
some degeneracies of the spectral values, it is also unique \cite%
{Beltrametti}. The condition 
\begin{equation}
\hat{\rho}_{s}\left( t\right) =\sum_{k}w_{k}\hat{\rho}_{k}\left( t\right) ,
\end{equation}%
for such a decomposition can be fulfilled in many different ways, but in
general the states appearing in this decomposition are not accessible by any
measurement (see e.g. \cite{Schroeck}). On the other hand, the measurement
procedure can be described by a $\sigma $-algebra $F\left( M\right) $ on
some measurable space $M$ \cite{Ludwig2}. In OQP, the instrument map $\cal
I$ is a $\sigma$-additive set function defined on $F(M)$ with values in
the set of positive operators on ${\cal H}_{s}$ (see \cite{Ludwig2}). For any
set $X\in F(M)$, the instrument map ${\cal I}\left( X\right) $ is an 
operation in Davies' sense \cite{Davies}.

\section{\protect\bigskip Description of the non-ideal measurement}

In the usual formulation of quantum mechanics, the measurements are defined
by observables and their spectral decomposition. This description is
appropriate for an ideal measurement process (see, for example, 
\cite{Ludwig,Davies,Ludwig2} and references therein). For non-ideal
measurements, the relevant object is the instrument, which is introduced
in \cite{Ludwig2} (see also \cite{Schroeck}).

For a non-ideal measurement process described by the instrument $\cal I$, one
can define the following quantities: the probability for the result in the
set $X\in F\left( M\right) $ and the conditional state $\hat{\rho}%
_{X} $ in the case when the measurement result belongs to $X$: 
\begin{equation}
p\left( X\right) =\text{Tr}_{s}\left[ {\cal I}\left( X\right) \hat{\rho}_{s}
\right] ,
\end{equation}%
\begin{equation}
\hat{\rho}_{X}=\frac{{\cal I}\left( X\right) \hat{\rho}_{s}}{\text{Tr}_{s}%
\left[ {\cal I}\left( X\right) \hat{\rho}_{s}\right] }.
\end{equation}%
The set of operators 
\begin{equation}
\widehat{F}\left( X\right) ={\cal I}^{\ast }\left( X\right) \left( \hat{1}%
\right) 
\end{equation}%
is called the effect-valued measure (EVM) associated to the instrument $\cal
I$. Here ${\cal I}^{\ast }\left( X\right) $ is the dual of $\cal I$, acting
on the set of bounded operators on ${\cal H}_{s}$, ${\cal I}^{\ast }\left(
X\right) :{\cal L}\left( {\cal H}_{s}\right) \rightarrow {\cal L}\left( 
{\cal H}_{s}\right) $. The operators $\widehat{F}\left( X\right) $ are
selfadjoint, positive and
\begin{equation}
\widehat{F}\left( \emptyset \right) =0,\quad \widehat{F}\left( M\right) =%
\widehat{1}.
\end{equation}%
They are called effects. The EVM is $\sigma$-additive and the map 
$F\left( M\right) \ni X\rightarrow \widehat{F}\left( X\right) $ is called a
positive operator valued measure (POVM). For details see \cite{Davies,Ludwig2}%
.

In the Stern--Gerlach experiment considered here, the observable is the spin
projection on a fixed direction $O_{z}$. Denoting by $\widehat{O}_{z}$ the
corresponding operator, with eigenstates 
\begin{equation}
\widehat{O}_{z}\left| \uparrow \right\rangle =+\left| \uparrow \right\rangle
,\quad \widehat{O}_{z}\left| \downarrow \right\rangle =-\left| \downarrow
\right\rangle ,
\end{equation}%
we will assume that the EVM associated to the instrument $\cal I$ has the form
\begin{eqnarray}
\widehat{F}\left( \left\{ \uparrow \right\} \right) &=&\widehat{F}_{\uparrow
}=\frac{1}{2}\left( \hat{1}+\vec{\xi}\cdot \overrightarrow{\hat{\sigma}}%
\right) , \\
\widehat{F}\left( \left\{ \downarrow \right\} \right) &=&\widehat{F}%
_{\downarrow }=\frac{1}{2}\left( \hat{1}-\vec{\xi}\cdot \overrightarrow{\hat{%
\sigma}}\right) .
\end{eqnarray}%
Here $\vec{\xi}$ is a three dimensional real vector, with $\left\vert \vec{\xi%
}\right\vert \leq 1$, $\overrightarrow{\hat{\sigma}}=\left( \widehat{\sigma}%
_{x},\widehat{\sigma}_{y},\widehat{\sigma}_{z}\right) $ and $\widehat{\sigma}%
_{i}$ are the Pauli matrices; `$\uparrow $' and `$\downarrow $' denote the
two possible measurement results. For $\left\vert \vec{\xi}\right\vert =1$,
the effects are projectors, and one recovers the ideal measurement.

The instrument describing the Stern--Gerlach experiment is assumed to be of
Kraus form (see \cite{Kraus1,Ludwig2}). For two possible results, the
instrument is determined by two operators $\widehat{A}_{\uparrow }$ and $%
\widehat{A}_{\downarrow }$: 
\begin{equation}
{\cal I}\left( \left\{ \uparrow \right\} \right) \hat{\rho}_{s}=\widehat{A}%
_{\uparrow }\hat{\rho}_{s}\widehat{A}_{\uparrow }^{\dagger },
\end{equation}%
\begin{equation}
{\cal I}\left( \left\{ \downarrow \right\} \right) \hat{\rho}_{s}=\widehat{A}%
_{\downarrow }\hat{\rho}_{s}\widehat{A}_{\downarrow }^{\dagger }.
\end{equation}%
These operators must satisfy
\begin{equation}
\widehat{A}_{\uparrow }^{\dagger }\widehat{A}_{\uparrow }+\widehat{A}%
_{\downarrow }^{\dagger }\widehat{A}_{\downarrow }=\widehat{1}
\end{equation}%
and
\begin{equation}
\widehat{F}_{\uparrow }=\widehat{A}_{\uparrow }^{\dagger }\widehat{A}%
_{\uparrow },\quad \widehat{F}_{\downarrow }=\widehat{A}_{\downarrow
}^{\dagger }\widehat{A}_{\downarrow }.
\end{equation}

In the following, we will work in the Bloch representation. An arbitrary
state can be written as 
\begin{equation}
\hat{\rho}_{s}=\frac{1}{2}\left( \widehat{1}+\vec{k}\cdot \overrightarrow{%
\widehat{\sigma}}\right) ,
\end{equation}%
with 
\begin{equation}
\left\vert \vec{k}\right\vert \leq 1.
\end{equation}%
The operators $\widehat{A}_{\uparrow }$ and $\widehat{A}_{\downarrow }$ can
be written as 
\begin{equation}
\widehat{A}_{\uparrow (\downarrow )}=\alpha _{\uparrow (\downarrow )}\hat{1}+%
\vec{\beta}_{\uparrow (\downarrow )}\cdot \overrightarrow{\widehat{\sigma}}, 
\end{equation}%
where $\alpha _{\uparrow (\downarrow )}$ are complex numbers and $\vec{\beta}%
_{\uparrow (\downarrow )}$ complex three dimensional vectors. In this
representation, the completeness condition becomes
\begin{equation}
\sum_{i=\uparrow ,\downarrow }\left( \left\vert \alpha _{i}\right\vert
^{2}+\left\vert \vec{\beta}_{i}\right\vert ^{2}\right) =1
\end{equation}%
and
\begin{equation}
\sum_{i=\uparrow ,\downarrow }\left( \alpha _{i}\vec{\beta}_{i}^{\ast
}+\alpha _{i}^{\ast }\vec{\beta}_{i}+i\vec{\beta}_{i}\times \vec{\beta}%
_{i}^{\ast }\right) =0.
\end{equation}%
On the other hand, from the definition of the EVM, one can see that
\begin{equation}
\widehat{F}_{i}=\widehat{A}^{\dagger }_{i}\widehat{A}_{i}.
\end{equation}%
Using Eq. (17) and the identity 
\begin{equation}
\left( \vec{a}\cdot \vec{\sigma}\right) \left( \vec{b}\cdot \vec{\sigma}%
\right) =\left( \vec{a}\cdot \vec{b}\right) \hat{1}+i\left( \vec{a}\times 
\vec{b}\right) \cdot \vec{\sigma}
\end{equation}%
we obtain 
\begin{equation}
\widehat{A}^{\dagger }_{i}\widehat{A}_{i}=\left( \left\vert \alpha
_{i}\right\vert ^{2}+\left\vert \vec{\beta}_{i}\right\vert ^{2}\right) \hat{1%
}+\vec{\xi}_{i}\cdot \overrightarrow{\widehat{\sigma}},
\end{equation}%
where 
\begin{equation}
\vec{\xi}_{i}=\alpha _{i}\vec{\beta}_{i}^{\ast }+\alpha _{i}^{\ast }\vec{%
\beta}_{i}+i\vec{\beta}_{i}\times \vec{\beta}_{i}^{\ast }.
\end{equation}%
One can see that $\vec{\xi}$ in Eq. (7) is given by 
\begin{equation}
\vec{\xi}=\vec{\xi}_{\uparrow }-\vec{\xi}_{\downarrow }.
\end{equation}

The purpose of the next section is to determine the parameters $\alpha
_{\uparrow (\downarrow )}$ and $\vec{\beta}_{\uparrow (\downarrow )}$ from
experimentally accessible quantities, by using the probabilities of single
and successive measurements.

\section{Results for the small non-ideality case}

The general problem as formulated above is very difficult to solve because
of the quadratic character of the relations between the parameters $\alpha
_{i}$, $\vec{\beta}_{i}$ and the experimentally accessible quantities. In
order to get some explicit formulae, one can consider a small non-ideality
case, by assuming that the measurement device is almost ideal, so that it
can be considered as a small perturbation of an ideal Stern--Gerlach filter.

Let us consider first a single measurement. The probability to obtain the
result `$\uparrow$' when the state is $\hat{\rho}_{s}$ is 
\begin{equation}
f\left( \vec{k}\right) _{\uparrow }=\text{Tr}\left[ \hat{\rho}_{s}\widehat{F}%
_{\uparrow }\right] =\frac{1}{2}\left( 1+\vec{k}\cdot \vec{\xi}\right) .
\end{equation}%
The deviation from the ideal case is given by 
\begin{equation}
\delta f\left( \vec{k}\right) _{\uparrow }=f\left( \vec{k}\right) _{\uparrow
}-f_{0}\left( \vec{k}\right) _{\uparrow },
\end{equation}%
where $f_{0}\left( \vec{k}\right) _{\uparrow }$ is the ideal case
probability. In order to get explicit relations, it is convenient to write
the vector $\vec{k}$ in spherical coordinates: 
\begin{equation}
\vec{k}=(\sin \theta \cos \varphi ,\sin \theta \sin \varphi ,\cos \theta ).
\end{equation}%
In the case of small non-ideality, one can consider the following
parametrization:
\begin{equation}
\alpha _{\uparrow (\downarrow )}=\frac{1}{2}+\eta a_{\uparrow (\downarrow )},
\end{equation}%
\begin{equation}
\vec{\beta}_{\uparrow (\downarrow )}=\pm \frac{1}{2}\vec{e}_{z}+\eta \vec{b}%
_{\uparrow (\downarrow )},
\end{equation}%
where $\eta $ is a small real parameter, $a_{\uparrow (\downarrow )}$ 
complex numbers and $\vec{b}_{\uparrow (\downarrow )}$ complex three
dimensional vectors. In order to simplify the following formulae, it is
convenient to write 
\begin{equation}
a_{\uparrow (\downarrow )}=a_{r\uparrow (\downarrow )}+ia_{i\uparrow
(\downarrow )},
\end{equation}%
\begin{equation}
\vec{b}_{\uparrow (\downarrow )}=\vec{b}_{r\uparrow (\downarrow )}+i\vec{b}%
_{i\uparrow (\downarrow )}.
\end{equation}%
The indices $r$ and $i$ stand for real and imaginary parts, respectively.

Using the above parametrization and keeping the terms up to first order in $%
\eta $, after some calculation one obtains for the deviation $\delta f\left( 
\vec{k}\right) _{\uparrow }$: 
\begin{eqnarray}
\delta f\left( \vec{k}\right) _{\uparrow } &=&\left[ a_{r}+b_{rz}+ \right. 
\nonumber \\
&&\left. \left( a_{r}+b_{rz}\right) \sin \theta \cos \varphi +\left(
b_{rx}-b_{iy}\right) \sin \theta \sin \varphi \right. \nonumber \\
&&\left. +\left( b_{ry}+b_{ix}\right) \cos \theta \right] _{\uparrow }.
\end{eqnarray}%
This expression can be written in the form 
\begin{equation}
\delta f\left( \vec{k}\right) _{\uparrow }=\left[
c_{0}+c_{1}\sin \theta \cos \varphi +c_{2}\sin \theta \sin \varphi
+c_{3}\cos \theta \right] _{\uparrow },
\end{equation}%
where $c_{0}$, $c_{1}$, $c_{2}$ and $c_{3}$ are real parameters. These
parameters can be determined from the fit of the experimental data. 

From Eqs. (37)--(42), one finds 
\begin{equation}
a_{r\uparrow }+b_{rz\uparrow }=c_{0\uparrow },
\end{equation}%
\begin{equation}
\left( a_{r}+b_{rz}\right) _{\uparrow }-\left( a_{r}+b_{rz}\right)
_{\downarrow }=2c_{1\uparrow },
\end{equation}%
\begin{equation}
\left( b_{rx}-b_{iy}\right) _{\uparrow }-\left( b_{rx}-b_{iy}\right)
_{\downarrow }=2c_{2\uparrow },
\end{equation}%
\begin{equation}
\left( b_{ry}+b_{ix}\right) _{\uparrow }-\left( b_{ry}+b_{ix}\right)
_{\downarrow }=2c_{3\uparrow }.
\end{equation}%
The same procedure can be applied for the result `$\downarrow $', so that
the deviations $\delta f\left( \vec{k}\right) _{\downarrow }$ are also
expressed in terms of parameters $c_{0\downarrow }$, $c_{1\downarrow }$, $%
c_{2\downarrow }$ and $c_{3\downarrow }$, and, similarly, one finds
\begin{equation}
a_{r\downarrow }+b_{rz\downarrow }=c_{0\downarrow },
\end{equation}%
\begin{equation}
\left( a_{r}+b_{rz}\right) _{\uparrow }+\left( a_{r}+b_{rz}\right)
_{\downarrow }=2c_{1z\uparrow },
\end{equation}%
\begin{equation}
\left( b_{rx}-b_{iy}\right) _{\uparrow }+\left( b_{rx}-b_{iy}\right)
_{\downarrow }=2c_{1y\uparrow },
\end{equation}%
\begin{equation}
\left( b_{ry}+b_{ix}\right) _{\uparrow }+\left( b_{ry}+b_{ix}\right)
_{\downarrow }=2c_{1x\uparrow }.
\end{equation}%

For a successive measurement, when the second device is rotated with respect
to the first, a similar analysis can be done. In this case, the parameters
$c_{1z\uparrow }$, $c_{1y\uparrow }$ and $c_{1x\uparrow }$ are specific to each
rotation, and can be calculated from the corresponding fit of the
experimental data. Then, the following equations are obtained:
\begin{equation}
\left( a_{r}+b_{rz}\right) _{\uparrow }+\left( a_{r}+b_{rz}\right)
_{\downarrow }=2c_{1z\uparrow },
\end{equation}%
\begin{equation}
\left( b_{rx}-b_{iy}\right) _{\uparrow }+\left( b_{rx}-b_{iy}\right)
_{\downarrow }=2c_{1y\uparrow },
\end{equation}%
\begin{equation}
\left( b_{ry}+b_{ix}\right) _{\uparrow }+\left( b_{ry}+b_{ix}\right)
_{\downarrow }=2c_{1x\uparrow },
\end{equation}%
where the second index from the experimental parameters stays for the main
axis of the device after the rotation (\ref{sgrotvect}), which, again, is
expected to be sufficient for the description of the two-step successive
measurements. For more-step successive measurements one can proceed in the
same manner.

\section{Conclusion}

In this work we have described a real, non-ideal Stern--Gerlach apparatus as a
two-outcome quantum instrument in the framework of operational quantum physics.
For a spin-1/2 system, each outcome of the filter measurement is represented by
a Kraus operator of the form
\begin{equation}
  A_{\uparrow(\downarrow)} = \alpha_{\uparrow(\downarrow)} \mathbf{1}
  + \vec{\beta}_{\uparrow(\downarrow)} \cdot \vec{\sigma},
\end{equation}
subject to the usual completeness relation. The corresponding effects
\(
  F_{\uparrow(\downarrow)} = A_{\uparrow(\downarrow)}^{\dagger}
  A_{\uparrow(\downarrow)}
\)
generalize the ideal projectors associated with measurement of the spin
component along the $z$ axis.

We have shown that the probabilities of single and successive measurements, for
arbitrary preparation directions and for a finite set of rotations of the
apparatus, can be written explicitly in terms of these parameters. In the
general case this leads to a system of quadratic equations, which is difficult
to solve explicitly. However, when the deviation from an ideal Stern--Gerlach
filter is small, one can expand around the projective case and work to first
order in a small non-ideality parameter. In this perturbative regime the
equations become linear, and the deviations of the observed probabilities from
the ideal values can be fitted by simple angular functions on the Bloch sphere.
The fitted coefficients then determine the parameters entering the effects and,
when combined with data from two-step measurements with rotated instruments,
also constrain the remaining instrument parameters.

In this way, the operational formalism yields experimentally accessible
quantities (the coefficients of the angular fits) which can be directly related
to the parameters of a non-ideal Stern--Gerlach instrument. The compatibility
between the measured deviations and the linear relations derived here
constitutes a concrete test of the OQP description of non-ideal spin
measurements. Although we have restricted ourselves to spin-1/2 systems and to
small non-idealities, the method can in principle be extended to more strongly
non-ideal devices (by addressing the full quadratic system) and to longer
sequences of measurements. These extensions, as well as a more detailed
analysis of specific physical sources of non-ideality, are left for future
work.
\bibliographystyle{apsrev4-2}
\bibliography{sg_2025}

@book{Ludwig,
  author    = {G. Ludwig},
  title     = {Foundations of Quantum Mechanics},
  publisher = {Springer},
  address   = {Berlin},
  year      = {1983},
}

@book{Davies,
  author    = {E. Davies},
  title     = {Quantum Theory of Open Systems},
  publisher = {Academic},
  address   = {London},
  year      = {1976},
}

@book{Ludwig2,
  author    = {G. Ludwig},
  title     = {An Axiomatic Basis for Quantum Mechanics},
  publisher = {Springer},
  address   = {Berlin},
  year      = {1985},
}

@book{Ludwig3,
  author    = {G. Ludwig},
  title     = {Wave Mechanics},
  publisher = {Springer},
  address   = {Berlin},
  year      = {1987},
}

@book{Kraus1,
  author    = {K. Kraus},
  title     = {States, Effects and Operations},
  publisher = {Springer},
  address   = {Berlin},
  year      = {1983},
}

@article{Martinez,
  author  = {J. Martinez},
  title   = {},
  journal = {Found. Phys.},
  volume  = {18},
  year    = {1988},
  pages   = {159},
}

@article{Lahti,
  author  = {P. Lahti},
  title   = {},
  journal = {Found. Phys.},
  volume  = {19},
  year    = {1989},
  pages   = {351},
}

@article{Busch,
  author  = {P. Busch},
  title   = {},
  journal = {Found. Phys.},
  volume  = {19},
  year    = {1989},
  pages   = {1257},
}

@book{Busch1,
  author    = {P. Busch and M. Grabowski and P. Lahti},
  title     = {Operational Quantum Physics},
  publisher = {Springer},
  address   = {Berlin},
  year      = {1995},
}

@book{Schroeck,
  author    = {F. Schroek},
  title     = {Quantum Mechanics on Phase Space},
  publisher = {Kluwer},
  address   = {Dordrecht},
  year      = {1996},
}

@book{Beltrametti,
  author    = {E. Beltrametti and G. Cassinelli},
  title     = {The Logic of Quantum Mechanics},
  publisher = {Addison-Wesley},
  address   = {Reading, Massachusetts},
  year      = {1981},
}

\end{document}